\documentclass[twocolumn]{aastex631}
\usepackage{graphicx}
\usepackage{dcolumn}
\usepackage{bm}
\usepackage{amssymb}
\usepackage{amsmath}

\begin{document}

\title{High-energy Neutrino Emission Associated with GWs from Binary Black Hole Mergers in AGN Accretion Discs}

\author{Zi-Hang Zhou}
\affiliation{Department of Astronomy, School of Physics, Huazhong University of Science and Technology, Wuhan 430074, China}

\author[0000-0003-4976-4098]{Kai Wang}
\affiliation{Department of Astronomy, School of Physics, Huazhong University of Science and Technology, Wuhan 430074, China}

\correspondingauthor{Kai Wang}
\email{kaiwang@hust.edu.cn}

\begin{abstract}
The search for multi-messenger signals of binary black hole (BBH) mergers is crucial to understanding the merger process of BBH and the relative astrophysical environment. Considering BBH mergers occurring in the active galactic nuclei (AGN) accretion disks, we focus on the accompanying high-energy neutrino production from the interaction between the jet launched by the post-merger remnant BH and disk materials. Particles can be accelerated by the shocks generated from the jet-disk interaction and subsequently interact with the disk gas and radiations to produce high-energy neutrinos through hadronic processes. We demonstrate that the identification of the high-energy neutrino signal from BBH merger in AGN disks is feasible. In addition, the joint BBH gravitational wave (GW) and neutrino detection rate is derived, which can be used to constrain the BBH merger rate and the accretion rate of the remnant BH based on the future associated detections of GWs and neutrinos. To date, an upper limit of BBH merger rate density in AGN disks of $R_0 \lesssim 3\,\rm Gpc^{-3} yr^{-1}$ is derived for the fiducial parameter values based on the current null association of GWs and neutrinos.

\end{abstract}

\keywords{Neutrino Astronomy (1100); Active galactic nuclei (16); Black holes (162); Gravitational waves (678); High energy astrophysics (739)}

\section{Introduction} \label{sec:intro}
Gravitational waves (GWs) produced by stellar-mass binary black hole (BBH) mergers have been prime targets for Earth-based GW detectors, including Advanced LIGO~\citep{Aasi_2015}, Advance Virgo~\citep{Acernese_2015}, KAGRA~\citep{PhysRevD.88.043007}. The BBH system can be formed from an isolated stellar binary origin (e.g., the stellar evolution of binary massive stars~\citep{2002ApJ...572..407B,2012ApJ...759...52D,2016MNRAS.458.2634M}) or dynamical interactions in dense stellar systems, e.g., in globular clusters~\citep{PhysRevD.93.084029,2016ApJ...824L...8R}, in quiescent galactic nuclei~\citep{2014ApJ...794..106A,Antonini_2016,10.1093/mnras/stz1803} or the Active galactic nuclei (AGN) accretion disks~\citep{1993ApJ...409..592A,Tagawa_2020,2023ApJ...944..159F,li2023hydrodynamical}.

Especially, AGN accretion disk can be a promising factory of compact-stellar binaries (CNBs) including BBH systems since it generally contains various stars and compact objects, such as white dwarfs (WDs), neutron stars (NSs), and stellar mass 
black holes (BHs). Due to the high density of AGN accretion disk, many stars and compact objects may be captured from nuclear star clusters or migrate from the outer self-gravitating region via gravitational instability \citep{syer1991star,artymowicz1993star,kolykhalov1980outer,goodman2004supermassive}. Due to the rich number of stars and compact objects in the accretion disk, there can be a high probability of forming binary systems, including binary neutron stars (BNSs), neutron star–black holes (NSBHs), and BBHs~\citep{bartos2017rapid,leigh2018rate}.

The simultaneous detections of GW and electromagnetic (EM) signals have been long expected. Unlike BNS and NSBH mergers, the BBH mergers are usually believed not to generate EM radiations due to a lack of accretion materials to power a jet. However, BBH mergers embedded in AGN disks can potentially power EM counterparts by accreting a significant amount of disk gas and then interacting with the disk medium~\citep{bartos2017rapid,mckernan2019,wang2021accretion,kimura2021outflow,tagawa2022}. In the observational aspect, EM follow-up observations have been implemented for LIGO/Virgo BBH merge events and seven potential AGN flares statistically associated with one or more of nine LIGO/Virgo events have been suggested~\citep{graham2023light}. In particular, the optical flare ZTF19abanrhr with a luminosity of $\sim 10^{45}\,\rm erg/s$ detected by the Zwicky Transient Facility (ZTF) in the AGN J124942.3+344929 was reported to be spatially coincidence with GW190521~\citep{graham2020candidate}, which is the heaviest BBH merger with a total mass of $M_{\mathrm{BBH}}\sim100M_{\odot}$ detected so far~\citep{PhysRevLett.125.101102}.

Although the joint observation of EM and GW signals is exciting and will help to better study the BBH merger process, it is still under debate whether the accompanying EM flare can be identified from the bright AGN disk emission. For instance, if the shock is too weak, the subsequent EM emission is too dim to be identified against bright AGNs and can be disclosed only for lower-luminosity AGNs~\citep{mckernan2019}. Besides, pre-merger outflows may also create a cavity around the BBH merger remnant to make the accretion rate insufficient to produce high luminosity~\citep{kimura2021outflow}.

In this \textit{letter}, we propose that the high-energy neutrino can be an alternative probe for the joint multi-messenger study for BBH merger embedded in AGN disks. Due to the negligible absorptions by AGN disk materials, the high-energy neutrinos can easily escape from the complicated AGN disk environment. The high-energy neutrino emission from AGN disks induced by some stellar activities therein has been explored~\citep{zhu2021thermonuclear,zhu2021high,Zhou_2023}.
However, the study of the high-energy neutrino production mechanism for BBH mergers insider AGN disks is still lacking although some searches of associated high-energy neutrino events have been explored~\citep{2023ApJ...944...80A}. In our scenario, the merged BH remnant is kicked out from the cavity and into the intact AGN disk environment and then accretes surrounding gas at the Bondi-Hoyle-Lyttleton rate \citep{kimura2021outflow,comerford2019bondi,wang2021accretion,graham2023light,tagawa2023observable}. The interaction between the post-merger Blandford-Znajek jet and the AGN disk medium can create various shocks, which in turn accelerates high-energy cosmic rays and produce neutrinos through hadronic processes. It is also expected to produce a detectable EM signal when shock breaks out from the disk~\citep{wang2021accretion,tagawa2023observable}. Therefore, the BBH merger inside AGN disks can be the potential triplet messenger (GW, EM, and neutrino) source.

\section{Jet Structure and Shock Acceleration} 

BBH mergers are expected to occur at about $r\sim10^3R_{\mathrm{g}}$ in the migration traps \citep{bellovary2016migration}, where $R_{\mathrm{g}}=GM_{\mathrm{SMBH}}/c^2$ is the Schwarzschild radius of central supermassive BH (SMBH), $G$ is the gravitational constant, $M_{\mathrm{SMBH}}$ is the SMBH mass, and $c$ is the speed of light. For a gas-pressure-dominated disk, the disk density is $\rho \left( h \right) =\rho _0\exp \left( -h/H \right)$, where $\rho_0$ is the density of mid-plane, $h$ is the vertical distance, and $H$ is the typical disk height. For a SMBH with mass $10^8M_{\odot}$ and the disk with aspect ratio $r/H\sim0.01$, one can get $H=1.5\times 10^{14}\mathrm{cm}$. The mid-plane density near the migration traps is approximately $10^{-10} \mathrm{g}\ \mathrm{cm}^{-3}$.

\begin{figure*}
	\centering
	\includegraphics[width = 0.99\linewidth , trim = 0 0 0 0,clip]{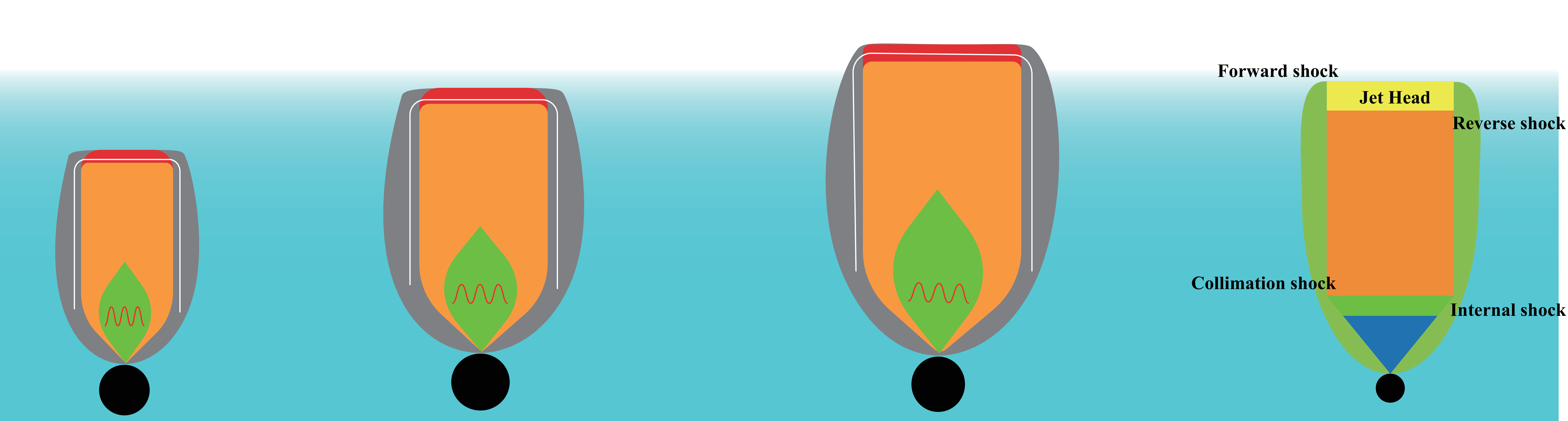}
	\caption{The three pictures on the left are schematic pictures of jet propagation in the AGN disk. There are four types of shocks in the jet, including forward shock (FS), reverse shock (RS), collimation shock (CS) and internal shock (IS). The picture on the far right is the schematic picture of an idealized jet used in our calculations. Each color corresponds to the same Lorentz factor in that region. The shocks are located at the junction of each shock area.}
	\label{fig:Jet}
\end{figure*}

A cocoon usually accompanies a jet in a dense environment. We consider a Blandford-Znajek jet launched from rapidly accreting and spinning remnant BH in an AGN disk. The Bondi-Hoyle-Lyttleton rate is adopted as \citep{comerford2019bondi,2021ApJ...911L..14W}
\begin{equation}
\dot{m}_{\mathrm{BHL}}=\frac{4\pi G^2M_{\mathrm{BBH}}^{2}\rho_0}{v_{\mathrm{rel}}^{3}}.
\end{equation}

Through simulation using a hierarchical population analysis framework, \citep{gayathri2023gravitational} considered binary formation in AGN disks along with phenomenological models and found that the high-mass and high-mass-ratio binaries appear more likely to have an AGN origin.
By referring to the GW event parameters \citep{gayathri2021black}, we adopt a total mass of $M_{\mathrm{BBH}}\sim100M_{\odot}$ for the BBH and assume that the relative velocity is given by $v_{\mathrm{rel}}=v_{\mathrm{k}}+c_s$. Here we use a gas sound speed of $c_s \sim 50\mathrm{km \, s^{-1}}$ and a kick velocity of $v_{\mathrm{k}}\sim 200 \, \mathrm{km \, s^{-1}}$ \citep{graham2020candidate}.
The jet kinetic luminosity $L_\mathrm{j}$ can be expressed as $L_{\mathrm{j}}=\eta_{\mathrm{j}}\dot{m}c^2$ with the jet conversion efficiency $\eta _{\mathrm{j}}$. The jet conversion efficiency exhibits a strong correlation with the BH spin and is approximately $\eta _{\mathrm{j}}\sim a_{\rm BH}^2$~\citep{tchekhovskoy2010black}, where $a_{\rm BH}$ is the BH dimensionless spin. Here $\eta _{\mathrm{j}}=0.5$ is adopted for a remnant BH spin $a_{\rm BH} \simeq 0.7$ consistent with observations~\citep{PhysRevLett.125.101102}. The accretion rate can be parameterized as $\dot{m}=f_{\mathrm{acc}}\dot{m}_{\mathrm{BHL}}$. We use $f_{\rm acc}=[0.1,1,10]$ respectively in our calculations, where a low $f_{\rm acc}$ for a jet might be anticipated as a result of winds (or wide-angle outflows) emanating from an accretion disk with a super-Eddington rate, while a high $f_{\rm acc}$ could potentially be attributed to the recoil kick of remnant BH so that more materials can be encountered in AGN disks~\citep{tagawa2023observable}. Then the jet kinetic luminosity can be written as
\begin{equation}
	L_{\mathrm{j}}=6.4\times10^{45} f_{\mathrm{acc}} \rho_{0,-10}\left( \frac{M_{\mathrm{BBH}}}{100M_{\odot}} \right)  \left( \frac{v_{\mathrm{rel}}}{250\mathrm{km} \, \mathrm{s}^{-1}} \right)  \mathrm{erg\, s}^{-1}
    \label{eq1}
\end{equation}
where $M_{\odot}$ is the solar mass and $\rho_{0,-10}=(\rho_0/10^{-10})\mathrm{g}\,\mathrm{cm}^{-3}$.

The jet is initially uncollimated because the pressure of the cocoon on the jet is not high enough, and it will become collimated as it travels. As suggested by \citep{bromberg2011propagation}, the jet will be collimated if $\left( \frac{L_j}{\bar{\rho}t^2\theta _{0}^{4}c^5} \right) ^{2/5}<\theta _{0}^{-4/3}$ while it is uncollimated if $\frac{L_j}{\bar{\rho}t^2\theta _{0}^{2}c^5}>\theta _{0}^{-4/3}$, where $\theta_0\approx0.2$ is the jet opening angle and $\bar{\rho}=\left( 1/R_h \right) \int_{2R_h}^{R_h}{\rho \left( h \right) dh}$ is the average density of the medium.

We study the high-energy neutrino emission from four possible sites, i.e., internal shocks, collimation shocks, forward shocks, and reverse shocks, formed during the relativistic jet of post-merger BH interacting with the disk atmosphere~\citep{yuan2020high}. Fig.~\ref{fig:Jet} schematically describes the evolution of the structure of the jet-cocoon system as well as the shocks inside the jet. Particle acceleration is driven by these shocks, and high-energy neutrinos are produced by $pp$ and $p\gamma$ interaction processes.

If the ambient medium is non-relativistic and the reverse shock is strong, the velocity of the jet head is \citep{bromberg2011propagation}
\begin{equation}
	\beta _h=\frac{\beta _j}{1+\tilde{L}^{-1/2}}.
\end{equation}
At the initial time $t_0$, one can get the initial height of the jet head $h_0\approx ct_0$, the initial average density of the disk $\bar{\rho} _0=\bar{\rho}\left( h_0 \right)$, the initial $\tilde{L}_0=\frac{L_j}{\rho _0t_{0}^{2}\theta _{0}^{2}c^5}$ and the initial jet head velocity $\beta _{h,0}=\frac{\beta _j}{1+\tilde{L_0}^{-1/2}}$. Therefore, the dynamic evolution process of the jet can be calculated with time. Noted that 
$\tilde{L}=\frac{L_j}{\bar{\rho}t^2\theta _{0}^{2}c^5}$ should be replaced with $\tilde{L}=\left( \frac{L_j}{\bar{\rho}t^2\theta _{0}^{4}c^5} \right) ^{2/5}$ when $\tilde{L}<\theta _{0}^{-4/3}$.
Through this method, we obtained the evolution of jet head height $R_{\mathrm{h}}=h$, $\tilde{L}$, jet head velocity $\beta_{\mathrm{h}}$ over the jet eruption time. The pressure of the cocoon in the collimated regime can be written as  $P_c\simeq t^{-4/5}L_{j}^{2/5}\bar{\rho}^{3/5}\theta _{0}^{2/5}$ \citep{bromberg2011propagation}, then the height of the collimation shock can be calculated by $R_{\mathrm{cs}}=2\sqrt{L_{\mathrm{j}}/cP_{\mathrm{c}}}$.

We calculate the structural evolution of the jet and find the jet becomes collimated at $100\,\mathrm{s}$ after the jet eruption, which is a quite short time compared with the duration of a BZ jet. At this moment, the height of the jet head now is $R_{\mathrm{h}}\sim10^{12}\mathrm{cm}\ll H$ for the fiducial $f_{acc}=1$. We assume the Lorentz factor of the unshocked material in the pre-collimation region to be $\Gamma_{\mathrm{j}}$. In this particular region, internal shocks emerge due to fluctuations in velocity within the outflow, resulting in the creation of gas shells exhibiting differential speeds. We may approximate the height of the internal shocks to be $R_{\mathrm{is}}=\min \left( R_{\mathrm{cs}},2\Gamma _{j}^{2}ct_{\mathrm{var}} \right)$ \citep{yuan2020high}, where $t_{\mathrm{var}}\simeq 1\,\mathrm{s}$ is the adopted variability timescale. Shock acceleration can be efficient only if the shock is collisionless. Therefore, we can obtain a radiation constraint on the upstream of the shock for efficient Fermi acceleration, which is described as \citep{murase2013tev}
\begin{equation}
	\tau _u=n_u\sigma _Tl_u\lesssim \min \left[ 1,0.1\Gamma _{\mathrm{rel}}/\left( 1+2\ln \Gamma _{\mathrm{rel}}^{2} \right) \right] 
	\label{eq2}
\end{equation}
where $\tau_u$ is the upstream optical depth, $n_u$ is the comoving number density of upstream material, $\sigma_T$ is the Thomson cross section, $l_u$ is the length scale of the upstream fluid, $\Gamma _{\mathrm{rel}}$ represents the relative Lorentz factor between the shock downstream and upstream. It means that an efficient particle acceleration can only occur when the shock has a sufficiently strong jump between the upstream and downstream.

To better represent the velocity relationship between different regions of the jet, we consider an idealized jet diagram in Fig.~\ref{fig:Jet} where the upstreams of the collimation shock and the reverse shock are downstream of the internal shock and the collimation shock, respectively. The jet head is downstream of the forward shock and the reverse shock. The comoving number density of the upstream of the collimation shock can be written as $n_{\mathrm{cs,}u}=L_{\mathrm{iso}}/(4\pi\Gamma_{\mathrm{rel}}^{2}R _{\mathrm{cs}}^{2}m_pc^3)$, where $L_{\mathrm{iso}}\approx2L_{\mathrm{j}}/\theta_0^2$ is the isotropic equivalent one-side jet luminosity and $m_p$ is the mass of proton. Here $\Gamma_{\mathrm{j}}\sim20$ we used is the Lorentz factor of the unshocked material. Upstream optical depth can be calculated as
\begin{equation}
	\tau _{\mathrm{cs},u}\approx\frac{L_{\mathrm{iso}}\sigma _T}{4\pi \Gamma _{\mathrm{j}}^{3}R_{\mathrm{cs}}m_pc^3}.
\end{equation}

The relative Lorentz factor of internal shock between the shock downstream and upstream is $\Gamma_{\mathrm{rel,is}}\approx\Gamma_{\mathrm{r}}/2\Gamma_{\mathrm{j}}\approx5$, and therefore upstream optical depth internal shock then can be derived by
\begin{equation}
	\tau _{\mathrm{is},u}\approx \frac{L_{\mathrm{iso}}\sigma _T}{4\pi \Gamma _{\mathrm{j}}^{3}\Gamma _{\mathrm{rel},\mathrm{is}}^{2}R_{\mathrm{is}}m_pc^3}.
\end{equation}
Similarly, one can get the upstream optical depth for the reverse shock as
\begin{equation}
	\tau _{\mathrm{rs},u}\approx \frac{L_{\mathrm{iso}}\sigma _TR_{\mathrm{h}}}{4\pi \Gamma _{\mathrm{j}}^{3}\Gamma _{\mathrm{rel},\mathrm{cs}}^{-2}R_{\mathrm{cs}}^{2}m_pc^3},
\end{equation}
where $\Gamma _{\mathrm{rel},\mathrm{cs}}\approx \Gamma _{\mathrm{j}}/2\Gamma _{\mathrm{j}1}\approx \Gamma _{\mathrm{j}}\theta _0/2\simeq 2$. For the forward shock, the upstream optical depth is
\begin{equation}
\tau _{\mathrm{fs},u}\approx \frac{\bar{\rho}\sigma _TR_{\mathrm{h}}}{m_p}\approx 6.24\times 10^3\bar{\rho}_{-10} \left( \frac{R_h}{H} \right) .
\end{equation}
Comparing with Equation~\ref{eq1}, we find that the forward shock is almost always inefficient in accelerating particles, resulting in inefficient neutrino production at the forward shock site, while the other three shocks always efficiently accelerate particles. Therefore, the neutrino production from forward shock is neglected in the next calculations.

\section{Neutrino Production} 
By the shock jump conditions \citep{piran1995hydrodynamic}, we can describe the internal energy and density evolution of the upstream and downstream shock by $e_{\mathrm{sh},\mathrm{d}}/n_{\rm sh,d}m_pc^2=\Gamma _{\mathrm{rel}}-1$ and $n_{\mathrm{sh},\mathrm{d}}/n_{\mathrm{sh},\mathrm{u}}\approx 4\Gamma _{\mathrm{rel}}$, where $n_{\mathrm{sh,u}}$ and $n_{\mathrm{sh,d}}$ are the comoving number density in the upstream and downstream shock respectively, $e_{\mathrm{sh}}$ is the comoving internal energy density in the shock. So the energy density of downstream of the shock is $e_{\mathrm{sh},\mathrm{d}}=4\Gamma _{\mathrm{rel}}(\Gamma _{\mathrm{rel}}-1)n_{\mathrm{sh},u}m_pc^2$. The photon temperature of the downstream of the shock is $k_BT=(15\hbar^3c^3\varepsilon_ee_{\mathrm{sh,d}}/\pi^2)^{1/4}$, where $\varepsilon_e \approx 0.1$ is the electron energy fraction \citep{zhu2021high}. During the next calculations, these thermal photons are treated as the background photon field for inverse Compton scattering (ICS), photomeson production $p\gamma$, and Bethe-Heitler processes. 

To calculate the neutrino emission efficiency, we need to estimate the cooling and acceleration timescales of the protons. The acceleration timescale is given by $t_{p,\rm acc}=\epsilon_p/eBc$, where $B=\sqrt{8\pi\varepsilon_Be_{\mathrm{sh,d}}}$ is the downstream magnetic field intensity and the magnetic field energy fraction is adopted as $\varepsilon_B = 0.1$. High energy protons cooling mainly include synchrotron radiation and ICS as radiative processes, $pp$, $p\gamma$, and Bethe-Heitler processes as hadronic interaction processes, and finally, the adiabatic process. 

For radiative processes, the cooling timescale of synchrotron radiation is
\begin{equation}
	t_{p,\rm syn}=\frac{6\pi m_{p}^{4}c^3}{\sigma _Tm_{e}^{2}B^2\epsilon _p}
\end{equation}
and ICS has the cooling timescale
\begin{equation}
	t_{p,\rm IC}=\begin{cases}
		\frac{3m_{p}^{4}c^3}{4\sigma _Tm_{e}^{2}n_{\gamma}\epsilon _{\gamma}\epsilon _p},&\epsilon _{\gamma}\epsilon _p<m_{p}^{2}c^4,\\
		\frac{3\epsilon _{\gamma}\epsilon _p}{4\sigma _Tm_{e}^{2}n_{\gamma}c^5}\,\,  ,&\epsilon _{\gamma}\epsilon _p>m_{p}^{2}c^4.\\
	\end{cases}
\end{equation}
where $\varepsilon_{\gamma}=2.7k_BT$ is the average thermal photon energy downstream and $n_{\gamma}=\varepsilon_ee_{\mathrm{sh,d}}/\varepsilon_{\gamma}$ is the average thermal photon density downstream.
For hadronic cooling mechanisms, the $pp$ scattering timescale is given by $t_{pp}=1/c\sigma_{pp}n_p\kappa_{pp}$, where $\sigma_{pp}$ \citep{kelner2006energy} is the cross section and $\kappa_{pp}\simeq0.5$ is the inelasticity. The cooling efficiency of $p\gamma$ process can be calculated by \citep{murase2007high}
\begin{equation}
	t_{p\gamma}^{-1}=\frac{c}{2\gamma _{p}^{2}}\int_{\bar{\epsilon}_{th,p\gamma}}^{\infty}{d\bar{\epsilon}}\sigma _{p\gamma}\left( \bar{\epsilon} \right) \kappa _{p\gamma}\left( \bar{\epsilon} \right) \bar{\epsilon}\int_{\bar{\epsilon}/2\gamma _p}^{\infty}{d\epsilon \epsilon ^{-2}\frac{dn}{d\epsilon}}
	\label{photon meson}
\end{equation}
where $\gamma_p=\epsilon_p/mc^2$, $\bar{\epsilon}$ is the photon energy in the rest frame of the proton and $\frac{dn}{d\epsilon}$ is the photon number density. $\bar{\epsilon}_{th,p\gamma}=145\mathrm{MeV}$ is the threshold energy for $p\gamma$ process, $\sigma_{p\gamma}$ and $\kappa_{p\gamma}$ represent the cross section \citep{kelner2008energy} and inelasticity \citep{stecker1968effect}, respectively. By replacing the cross section, inelasticity, and threshold energy in the Equation~\ref{photon meson} with those of the Bethe-Heitler process, one can get the cooling efficiency of the Bethe-Heitler process \citep{chodorowski1992reaction}. Finally, the adiabatic cooling timescale is $t_{p,\rm ad}=R_{\mathrm{sh}}/c\Gamma_{\mathrm{sh,d}}$.

Considering that only $pp$ and $p\gamma$ processes can produce high-energy neutrinos, the other processes suppress the production of neutrinos. We can write the proton suppression factor by involving various cooling processes as
\begin{equation}
	\zeta _{p,\rm sup}\left( \epsilon _{\nu _i} \right) =\frac{t_{p,pp}^{-1}+t_{p,p\gamma}^{-1}}{t_{p,pp}^{-1}+t_{p,p\gamma}^{-1}+t_{\rm BH}^{-1}+t_{p,\rm syn}^{-1}+t_{p,\rm IC}^{-1}+t_{p,\rm ad}^{-1}}.
\end{equation}

Besides, neutrinos are produced by the decay of pions and kaons created through $pp$ and $p\gamma$ processes, which will be suppressed by other cooling processes. The suppression factor of these mesons can be calculated by \citep{zhu2021high}
\begin{equation}
	\zeta _{i,\rm sup}\left( \epsilon _{\nu _i} \right) =\frac{t_{i,\rm dec}^{-1}}{t_{i,\rm dec}^{-1}+t_{i,\rm had}^{-1}+t_{i,\rm syn}^{-1}+t_{i,\rm IC}^{-1}+t_{i,\rm ad}^{-1}}
\end{equation}
where $i$ represents the meson produced by the $pp$ or $p\gamma$ processes.

Neutrino fluence for a single event can be obtained by calculating the summation of each neutrino channel by
\begin{equation}
	\epsilon _{\nu}^{2}F_{\nu}=\frac{1}{4\pi D_{L}^{2}}\sum_{i}{\int_{0}^{t_{\rm end}}{\frac{N_iL_{\rm iso}\zeta _{p,\rm sup}\left( \epsilon _{\nu _i} \right) \zeta _{i,\rm sup}\left( \epsilon _{\nu _i} \right)}{\ln \left( \epsilon _{p,\max}/\epsilon _{p,\min} \right)}dt}}
\end{equation}
where $N_{i}$ is the energy fraction that the jet energy converts to the neutrinos, and $i=$[$\pi,K,\mu _{\pi},\mu _K$] represents different neutrino production channels. Here $N_{\pi}=N_{\mu _{\pi}}=0.12$, $N_K=0.009$, and $N_{\mu _K}=0.003$. The neutrino energy is $\epsilon _{\nu _i}=a_i\epsilon _p$, where $a_{\pi}=a_{\mu_{\pi}}=0.05$, $a_K=0.1$ and $a_{\mu _K}=0.033$. We assume $t_{\rm end}=5\times10^6\,\rm s$ is the duration of the jet corresponding to the observational ZTF19abanrhr flare of GW190521, which lasts around tens of days~\citep{graham2020candidate}. $\epsilon _{p,\max}$ is the maximum proton energy calculated by $t_{\rm acc}=t_{\rm cool}$ and $\epsilon _{p,\min}\approx\Gamma_{\mathrm{sh,d}}m_p c^2$ is the minimum proton energy.

\begin{figure}
	\centering
	\includegraphics[width = 0.99\linewidth , trim = 15 0 15 10,clip]{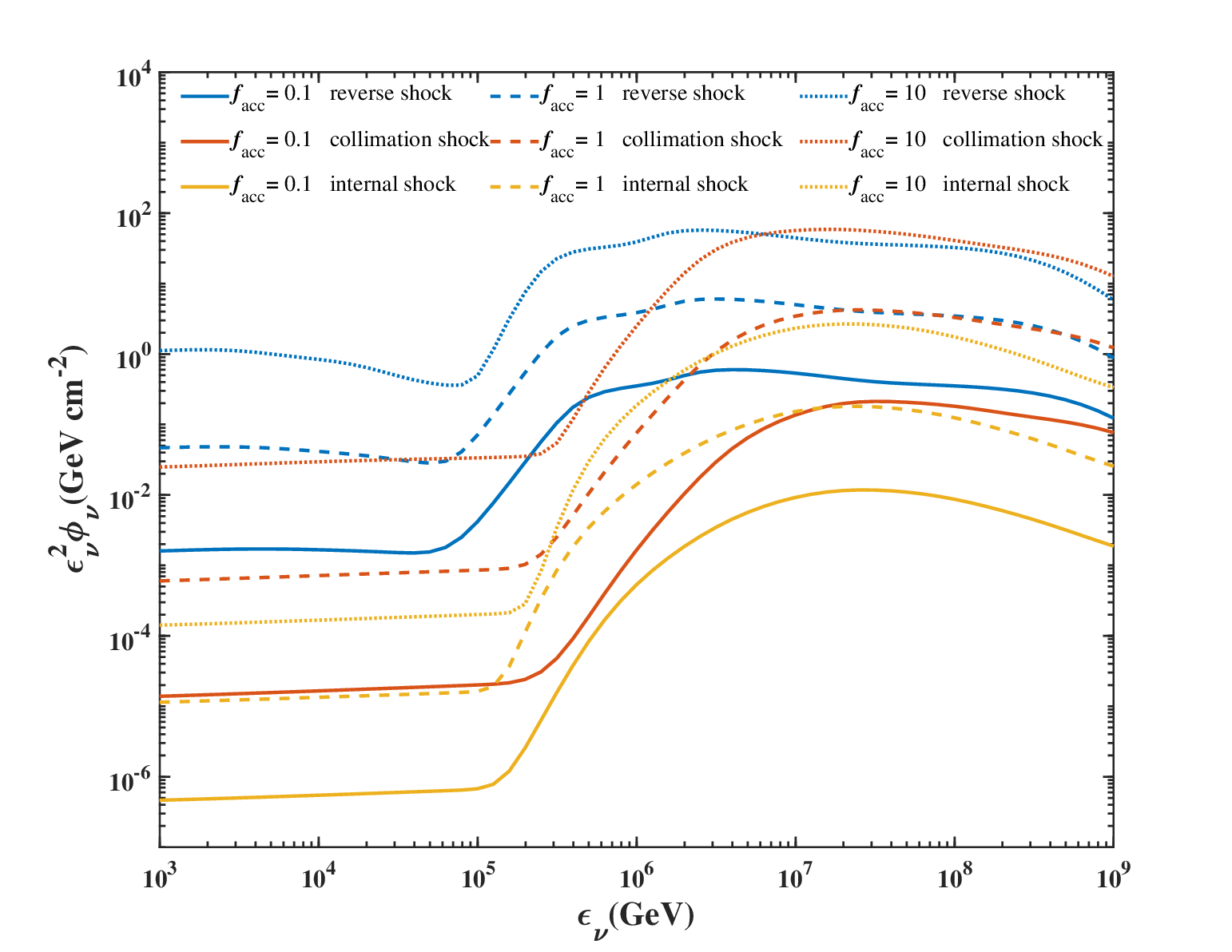}
	\caption{Neutrino fluences from diverse shocks at a distance of $100\,\rm Mpc$. The different colors represent different shock types. Each linetype represents different accretion rates, i.e., $0.1\dot{m}_{\mathrm{BHL}}$ (solid), $1\dot{m}_{\mathrm{BHL}}$ (dashed) to $10\dot{m}_{\mathrm{BHL}}$ (dotted).}
	\label{fig:Fluence}
\end{figure}

The produced neutrino fluence of a single BBH merger event occurred at $D_L=100\,\mathrm{Mpc}$ is shown in Figure \ref{fig:Fluence} including the reverse shock, collimation shock, and internal shock. We can see that the neutrino production of the collimation shock is comparable with that of the reverse shock above around few$\times 10^6\,\rm GeV$, while below this energy the reverse shock contributes more neutrino production. For three shock sites, at the lower energy part ($\lesssim 10^5\,\rm GeV$), all neutrino production suffers the significant suppression of the adiabatic cooling process, showing a lower fluence than the high-energy part. Note that here we consider the same jet duration $t_{\rm end}$ for different accretion rates, however, it may last a shorter time for higher accretion rate~\citep{wang2021accretion}.

\section{Neutrino and Joint GW+Neutrino Detection} 

\begin{figure}
	\centering
	\includegraphics[width = 0.99\linewidth , trim = 10 0 10 0,clip]{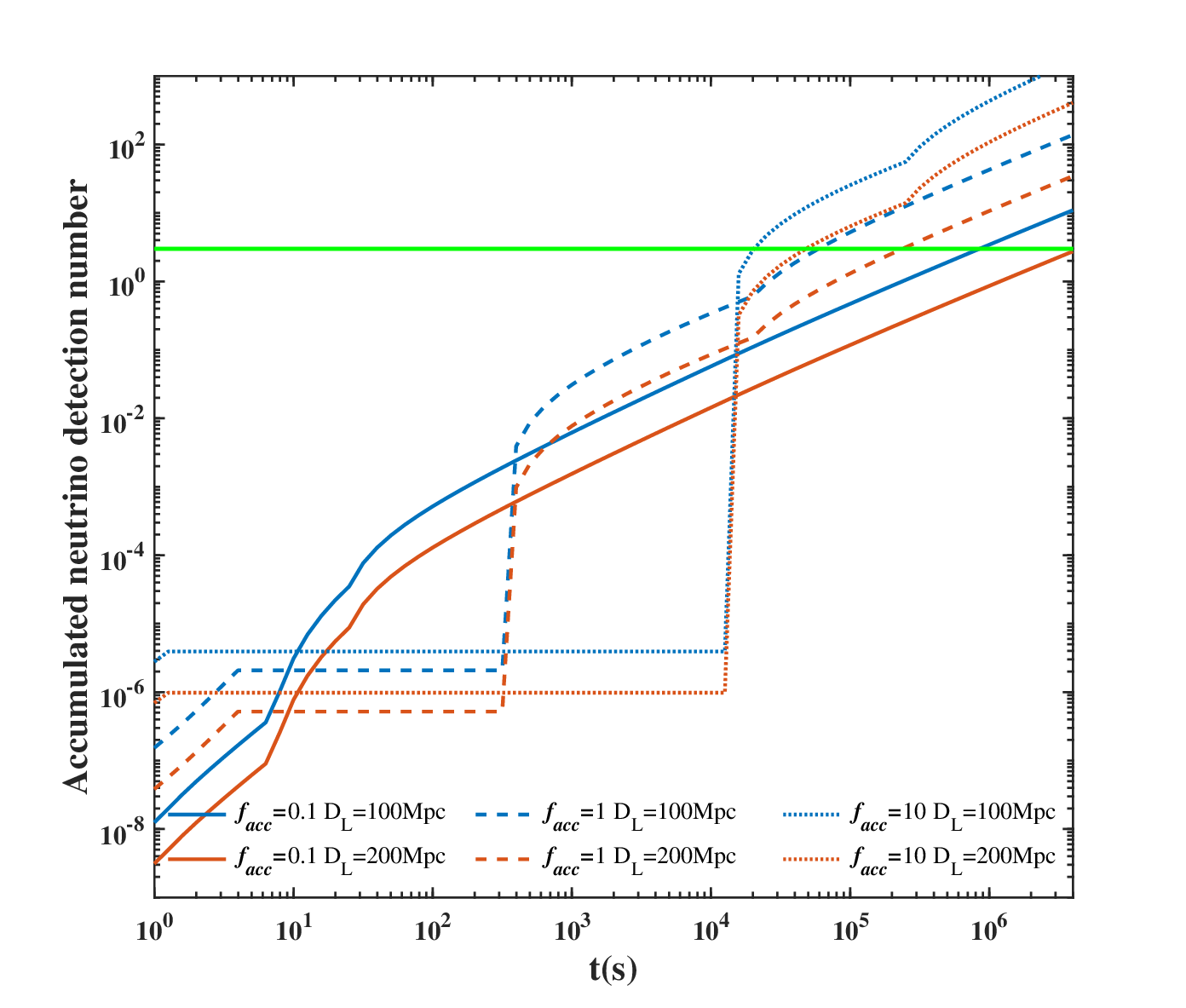}
	\caption{Cumulative detected neutrino number of a single source by IceCube with different accretion rates at two different locations, i.e., 100 Mpc and 200 Mpc. The green line indicates three detected neutrinos so that the neutrino detection by IceCube can be identified at $\sim 3 \sigma$ significance level.}
	\label{fig:cumulative neutrino}
\end{figure}

The all-flavor neutrino detection number can be calculated by

\begin{equation}
	N_{\nu}=\int{d\epsilon _{\nu}F_{\nu}\left( \nu _{\nu} \right) A_{\mathrm{eff}}\left( \epsilon _{\nu} \right)}
\end{equation}
where $A_{\mathrm{eff}}\left( \epsilon _{\nu _{\mu}} \right)$ is the effective area (100\,GeV--100\,PeV) of IceCube for a point source \citep{aartsen2020time}. The accumulative neutrino number with time is presented in Fig.~\ref{fig:cumulative neutrino}, it can be seen that the Ice-Cube can receive at least three neutrinos within $5\times10^6\,\rm s$ within a distance of $\lesssim 200\,\rm Mpc$. In a relatively optimistic situation ($f_{acc}=10$ and $D_{\mathrm{L}}=100\mathrm{Mpc}$), we can expect IceCube to receive three neutrinos within 6 hours which has reached a relatively high level of confidence. In the case of a high accretion rate ($f_{acc}=10$), the detection distance of the IceCube can reach 1 Gpc. However, as shown in Fig.~\ref{fig:cumulative neutrino}, for the high accretion rate, the particle acceleration tends to be forbidden at the early stage so that the neutrino production is low at the beginning.

\begin{table}    
 \centering 
    \footnotesize
	\caption{Joint BBH GW + Neutrino Detection Rate.}
    \vspace{0cm}
	\begin{tabular}{l@{\hskip 0.1in}l@{\hskip 0.1in}l}
 \hline
 \hline
	GW Detector & Detection Rate($\mathrm{yr}^{-1}$) & Detection Rate($\mathrm{yr}^{-1}$)\\
   							&	with IceCube &	with IceCube-Gen2	\\
 \hline
		aLIGO & $0.3210R_0$ & $1.0354R_0$\\
                  & (0.0007-6.423) & (0.0021-20.73)\\
 \hline  
		adVirgo & $0.3168R_0$ & $0.9575R_0$\\
                    & (0.0007-6.335) & (0.0019-19.15)\\
 \hline  
		KAGRA & $0.3168R_0$ & $0.9546R_0$\\
                     & (0.0006-6.335) & (0.0019-19.09)\\
 \hline  
		Voyager & $0.3240R_0$ & $1.0813R_0$\\
                     & (0.0007-6.480) & (0.0022-21.63)\\
 \hline  
		ET & $0.3254R_0$ & $1.0828R_0$\\
                     & (0.0007-6.508) & (0.0022-21.66)\\
 \hline
	\end{tabular}
	\label{table:joint}
 \footnotesize{{\leftline { $R_0\simeq [0.002,20]\,\mathrm{Gpc^{-3}\, yr^{-1}}$ is the BBH merger rate density} }}\\
 \footnotesize{{\leftline { within AGN disks~\citep{grobner2020binary}.} }}
\end{table}

\begin{figure}
	\centering
	\includegraphics[width = 0.99\linewidth , trim = 10 0 10 0,clip]{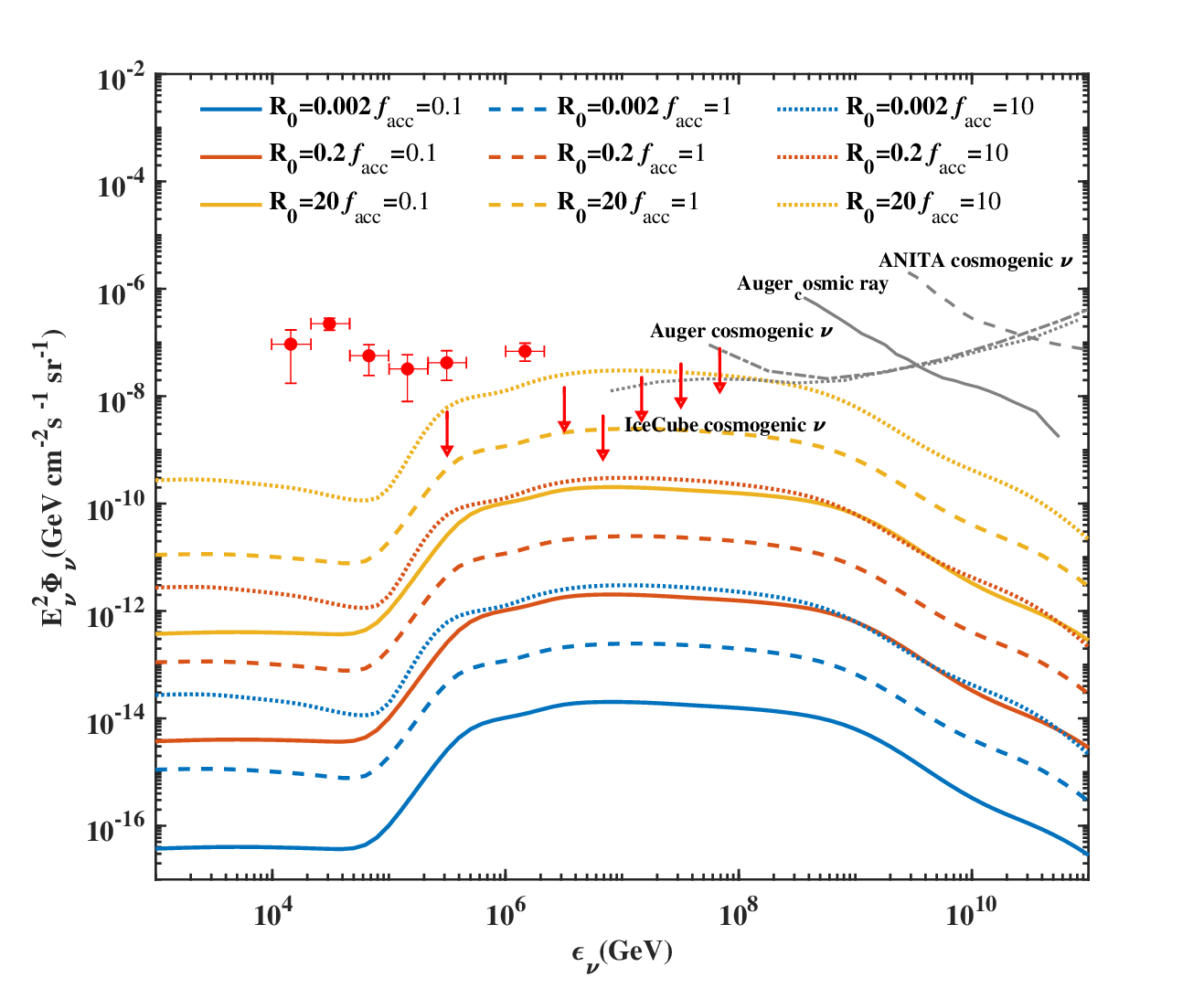}
	\caption{Expected all-flavor diffuse neutrino fluence contributed from BBH mergers in AGN disks. Diverse local event rates and accretion rates are considered. Red points and upper limits are the observed diffuse neutrino fluence by IceCube~\citep{aartsen2015combined,icecube2021detection}. The gray lines are 90\% upper limit of the cosmogenic neutrino for diverse instruments.}
	\label{fig:diff_fluence}
\end{figure}

We calculate the joint detection rate of neutrinos and GW by generating $3\times10^4$ random BBH merger events. $f_{acc}=1$ is used in the calculation. GW signal is treated as isotropic and high-energy neutrinos from the jet are beamed with a beaming correction factor $f_b=\theta_0^2/2$. As for the mass distribution of BBH, we adopt the results of \citep{gayathri2023gravitational}, which presented a one-parameter model for BBH formation and merger within an AGN disk and parameterized by the maximum mass $m_{\max}$ of the natal BH distribution. The mass distribution for $m_{\max}=75$ is adopted for calculations. We use Python's module PYCBC to generate the waveform of the GW signal and calculate the signal-to-noise ratio (SNR) of each event by 
$
\left( S/R \right) ^2=\int_{f_{\min}}^{f_{\max}}{\frac{|\tilde{h}\left( f \right) |^2}{S_n\left( f \right)}df}
$ 
\citep{zhu2021kilonova}. The threshold SNR=8 is involved to confirm the detection of GW signals, and the neutrino detection number $N_{\nu}=1$ is roughly employed as the threshold for confirming the detection of IceCube or IceCube-Gen2. The effective area of IceCube-Gen2 is taken as 6 times larger than IceCube's \citep{aartsen2021icecube}. The redshift evolution factor we adopted is $f\left( z \right) =\left[ \left( 1+z \right) ^{5.7\eta} +\left( \frac{1+z}{0.36} \right) ^{1.3\eta}+\left( \frac{1+z}{3.3} \right) ^{-9.5\eta}
+\left( \frac{1+z}{3.3} \right) ^{-24.5\eta} \right] ^{1/\eta}$~\citep{sun2015extragalactic}.
For multiple gravitational wave detectors, we calculated the joint detection rates separately. The detector sensitivities can be expressed as amplitude spectral densities (ASD, https://dcc.ligo.org/LIGO-T1500293/public). The result is shown in Table~\ref{table:joint}, where $R_0$ is the local BBH mergers event rate. It can be seen that although different GW detectors have different detection capabilities, the derived joint detection rates are largely identical with only minor differences for the same neutrino detector. This indicates that the detection ability of the neutrino detector is the main factor affecting the joint detection rate, and indeed, the joint detection rate significantly increases for IceCube-Gen2 with a larger effective area.

The diffuse neutrino fluence can be estimated by \citep{razzaque2004tev}
\begin{equation}
	\epsilon _{\nu,obs}^{2}\varPhi _{\nu}=\epsilon _{\nu ,obs}^{2}f_b\int_0^{z_{\max}}{dzR_0f\left( z \right) F_{\nu}\left( \epsilon _{\nu ,obs} \right) \frac{dV}{dz}}
\end{equation}
where $\epsilon_{\nu,\mathrm{obs}}=\epsilon_{\nu} /(1+z)$ is the observed neutrino energy $dV/dz=4\pi D_{\mathrm{L}}^{2}c/(1+z)^3/(H_0 \sqrt{\Omega_{\Lambda}+\Omega_{\mathrm{m}}(1+z)^3)}$ is the comoving volume. The standard $\mathrm{\Lambda CDM}$ cosmology $H_0=67.8 \mathrm{km\, s^{-1}}\mathrm{Mpc^{-1}}$,  $\Omega_{\Lambda}=0.692$ and $\Omega_{\mathrm{m}}=0.308$ is applied \citep{ade2016planck}. 
$R(z)=R_0f(z)$ is adopted as the redshift distribution of BBH mergers. The estimated GW rate density associated with BBH mergers lies in the range $R\sim(0.002-18)\,\mathrm{Gpc}^{-3}\mathrm{yr}^{-1}$ is given by \citep{grobner2020binary}. Recently, some literature gives the event rate to be $2.5 \,\mathrm{Gpc^{-3} yr^{-1}}$ or dozens $\mathrm{Gpc^{-3} yr^{-1}}$ \citep{gayathri2021black,gayathri2023gravitational}. Here three typical values of [0.002,0.2,20] as the possible local event rates are used. We sum the neutrino fluences of each type of shock and use the mass distribution of BBH mentioned above. Our result shows that BBH mergers in AGN disk contribute to the neutrino background at the relatively high energy above $10^6\,\rm GeV$, while they provide a relatively little contribution to the neutrino background at the low energy.

Note that the above joint detection rate of GWs and neutrinos in Table~\ref{table:joint} and the neutrino production are derived based on the adopted fiducial parameter values, e.g., the jet conversion efficiency $\eta_j=0.5$, the accretion factor $f_{\rm acc}=1$, and the jet eruption duration $t_{\rm end}=5 \times 10^{6}\,\rm s$. The jet conversion efficiency $\eta_j$ and the accretion factor $f_{\rm acc}$ are crucial quantities to determine the jet kinetic luminosity, which directly affects how many energies can be converted to neutrinos. Besides, for a specific jet luminosity, the jet eruption duration will significantly influence the final neutrino fluence as well. Although these values in our calculations have been calibrated by the observations for GW190521 and  ZTF19abanrhr optical flare~\citep{PhysRevLett.125.101102,graham2020candidate}, diverse BH mergers may have diverse values for these parameters, inducing a different conclusion. For example, although the jet eruption duration is involved with a comparable observational duration of ZTF19abanrhr flare, i.e., tens of days, the real jet eruption duration or accretion timescale may be shorter or even episodic~\citep{wang2021accretion}. For a shorter jet duration, e.g., $10^5\,\rm s$, the neutrino detection for a single BBH merger event by IceCube is feasible only for a higher accretion rate or a closer distance as shown in Fig.~\ref{fig:cumulative neutrino} and the joint GW + Neutrino detection rate will be lower since less neutrino production will be expected.

Another key parameter to affect joint GW + Neutrino detection rate is the BBH merger rate within AGN disks. However, its uncertainty is still quite large to date, inducing a large uncertainty on the final evaluation of joint GW + Neutrino detection rate as shown in Table~\ref{table:joint}. If the fiducial parameter values are involved, the high BBH merger rate density tends to be ruled out in order to be consistent with the current null association of GW and neutrino events recently reported in~\cite{2023ApJ...944...80A} and \cite{JointSearchingGiacomo2023}, resulting in an upper limit of BBH merger rate density in AGN disks of $R_0 \lesssim 3\,\rm Gpc^{-3} yr^{-1}$. The result greatly improves the early constraint on BBH merger rate in AGN disks~(e.g., $10^{-3}-10^4\,\rm Gpc^{-3} yr^{-1}$ in~\cite{2018ApJ...866...66M} and $0.02-60\,\rm Gpc^{-3} yr^{-1}$ in~\cite{2020ApJ...898...25T}). Comparing with the total BBH merger rate given by LIGO/Virgo, i.e., $R=53.2_{-28.8}^{+58.5}\,\rm Gpc^{-3} yr^{-1}$~\citep{2019ApJ...882L..24A}, BBH mergers in AGN disks will be small fraction of the total BBH mergers in the universe. However, the constraint can be alleviated if the real parameter values deviate from the fiducial values.

Moreover, based on our calculations of diffuse neutrino background from BBH mergers in AGN disks (see Fig.~\ref{fig:diff_fluence}), the most optimistic parameters, e.g., high BBH merger rate and high accretion rate of remnant BH at the same time, tend to be excluded due to the conflict with the diffuse neutrino observations.

\section{Discussions and Conclusions}
Neutrinos from remnant BH jet-induced shock acceleration make it possible to predict a BBH merger without the detection of an EM signal if the jet breakout brightness is overshadowed by the AGN accretion disk. In this letter, we consider a persistent BZ jet and investigate its structural evolution. Collimation shocks and reverse shocks contribute more neutrino production compared with internal shocks. The accumulative neutrino number over time is derived as well and we find that it is possible to receive enough neutrinos (exceeding three neutrinos) by IceCube within tens of days if the BBH mergers in AGN disk take place within a distance of a few hundred of Mpc and the remnant BH has an accretion rate of $f_{\rm acc} \gtrsim 0.1$ and other fiducial parameter values.

Based on 83 LIGO/Virgo BBH and lower-mass-gap merger alters~\citep{graham2023light} and the beaming correction, the source number with the neutrino detection by IceCube for these GW events can be estimated by $\sim 83 f_b\simeq 1.7 (\theta_0/0.2)^2$ if all BBH mergers occur in AGN disks and the neutrino emission from theses source can be identified. It implies only partial BBH mergers occurring in AGN disks or the neutrino emission from these sources can not be totally identified in order to be consistent with the current null association of GW and neutrino events~\citep{2023ApJ...944...80A}.

In addition, we calculate the joint GW + Neutrino detection rate by combining the diverse GW detectors and IceCube (or IceCube-Gen2). The uncertainty of joint detection is still quite large since the event rate of BBH merger in AGN disks is still quite unclear. However, the detection or non-detection of joint BBH GW + Neutrino association in the future can be used to constrain the BBH merger rate within AGN disks. For instance, the BBH merger rate in AGN disks should be lower if no association has been observed. A BBH merger rate density in AGN disks of $R_0 \lesssim 3\,\rm Gpc^{-3} yr^{-1}$ is derived for the fiducial parameter values based on the null association of GW and neutrino signals so far. In addition, the simultaneous high BBH merger rate in AGN disks and high accretion rate of remnant BH are excluded due to the observations of diffuse neutrino background.

BBH mergers in AGN disks are the ideal targets for multi-messenger observations and joint observations of EM, neutrino, and GW signals from them have caused more and more attention and have also been frequently explored recently. In the aspect of high-energy neutrinos, they can play an important role in the multi-messenger study of BBH mergers. The next-generation neutrino telescopes, e.g., IceCube-Gen2, Huge Underwater high-energy Neutrino Telescope (HUNT)~\citep{Huang:2023R8}, The tRopIcal DEep-sea Neutrino Telescope (TRIDENT)~\citep{2022arXiv220704519Y}, and the radio-Cherenkov neutrino detector ARIANNA~\citep{2020arXiv200409841A} and ARA~\citep{2019arXiv190711125A} combining with the next more advanced GW detectors can help to understand the nature of BBH mergers in AGN disks.

\begin{acknowledgments}
We thank Jin-Ping Zhu for the helpful discussions. We also thank the anonymous referee for the helpful comments, which have helped us to improve this paper. We acknowledge support from the National Natural Science Foundation of China under grant No.12003007 and the Fundamental Research Funds for the Central Universities (No. 2020kfyXJJS039).
\end{acknowledgments}

\vspace{5mm}

\bibliography{reference}{}
\bibliographystyle{aasjournal}

\end{document}